%% file: main.tex
\documentclass{jpp}
\usepackage{graphicx}

\usepackage[utf8]{inputenc}
\usepackage[T1]{fontenc}
\usepackage{amsmath}
\usepackage{physics}
\newcommand{\integral}{ \int_0 ^{2\pi}\!\!\!\!\!d\zeta \int_{0}^{2\pi}\!\!\!\!\!d\theta \int_{0}^1\!\!\! ds \,}
\usepackage[export]{adjustbox}

\usepackage{float}
\shorttitle{Coordinates constructions for toroidal domains}
\shortauthor{Z. Tecchiolli et al.}

\title{Constructing nested coordinates inside strongly shaped toroids using an action principle}

\author{Z. Tecchiolli\aff{1}
  \corresp{\email{zeno.tecchiolli@epfl.ch}},
  S. R. Hudson\aff{2},
  J. Loizu\aff{1}, 
  R. Köberl\aff{3},
  F. Hindenlang\aff{3}, 
  \and B. De Lucca\aff{1},
}

\affiliation{\aff{1}Ecole Polytechnique Federale de Lausanne (EPFL), Swiss Plasma Center (SPC), CH-1015 Lausanne, Switzerland \\
\aff{2} Princeton Plasma Physics Laboratory, PO Box 451, Princeton NJ 08543, USA \\
\aff{3}Max-Planck Institute for Plasma Physics, Garching, Germany
}

\begin{document}

\maketitle

\begin{abstract}
A new approach for constructing polar-like boundary-conforming coordinates inside a toroid with strongly shaped cross-sections is presented. A coordinate mapping is obtained through a variational approach, which involves identifying extremal points of a proposed action in the mapping space from $ [0, 2\pi]^2 \times [0, 1]$ to a toroidal domain in $\mathbb{R}^3$. This approach employs an action built on the squared Jacobian and radial length. Extensive testing is conducted on general toroidal boundaries using a global Fourier-Zernike basis via action minimization. The results demonstrate successful coordinate construction capable of accurately describing strongly shaped toroidal domains. The coordinate construction is successfully applied to the computation of 3D MHD equilibria in the GVEC code where the use of traditional coordinate construction by interpolation from the boundary failed. 
\end{abstract}


\input{Sections/Introduction}

\input{Sections/Mathematical_Formalism}

\input{Sections/Numerical_Implementation}


\input{Sections/Results}


\input{Sections/GVEC_application}


\input{Sections/Conclusions}

\input{Sections/Acknowledgments}
\input{Sections/Data}
\input{Sections/Interests}
\appendix

\input{Sections/Appendix}

\bibliographystyle{jpp}

\bibliography{jpp-instructions}

\end{document}

%% file: Sections/Introduction.tex
\section*{Introduction} \label{sec:whyweneedcoords}


The toroidal geometry, topologically characterised by a single hole and homeomorphic to a continuous surface formed by rotating a circle about an axis external to its plane, is fundamental in plasma physics due to its inherent capacity to provide closed magnetic flux surfaces, essential for confining and stabilising plasmas in  magnetic fusion devices. In that context,
any study of confinement, coil engineering, stability, and transport relies on the description of a magnetic field in a toroidal domain.

Using coordinates may not be the only possible choice to describe magnetic fields in toroidal geometry. For instance, some codes employ the surface integral method, such as BIEST \citep{o2018integral, malhotra2019taylor}, which represents the magnetic field using the generalised Debye source formulation. On the other hand, many different kinds of coordinate systems may be used, some of which are not boundary conforming. For example, M3D-C1 \citep{jardin2004triangular} exploits an unstructured cylindrical grid
because of the advantages of not having a coordinate singularity. Furthermore, cylindrical coordinates $(R, \phi, Z)$ are the natural choice for describing toroidal geometry owing to their close correspondence to a laboratory-frame and due to their simplicity in applying boundary conditions.

The construction of a polar-like coordinate grid, or coordinate mapping that is boundary-conforming, can be made explicit by the input boundary representing
the outermost surface of a set of non-intersecting nested torus-shaped surfaces, such that a monotonically increasing
radial coordinate exists with a starting point identifying the coordinates axis. This class of coordinate maps is very common in computations of magnetohydrodynamics (MHD) equilibria.

Given the double periodicity of toroidal boundaries, spectral methods \citep{matsushima1995spectral} are a popular choice.
These methods construct coordinates through a linear combination of global basis functions, expressing cylindrical coordinates $(R, \phi, Z)$ in terms of toroidal-like coordinates $(s, \theta, \zeta)$ via Fourier-polynomial decomposition. The main challenge resides in being able to correctly choose the coefficients for the basis functions in such a way that the mapping becomes bijective.

The challenges of such task are well depicted by codes routinely used for computing three-dimensional equilibria, in stellarators and tokamaks, such as VMEC \citep{hirshman1983steepest}, GVEC \citep{gvec_simons}, SPEC \citep{hudson2012computation, loizu2016verification}, or DESC \citep{dudt2020desc, conlin2023desc}. These codes solve the inverse problem of finding an equilibrium provided a choice for the flux surfaces.

\begin{figure}
\begin{minipage}{.5\textwidth}
\centering
\includegraphics[scale=0.36]{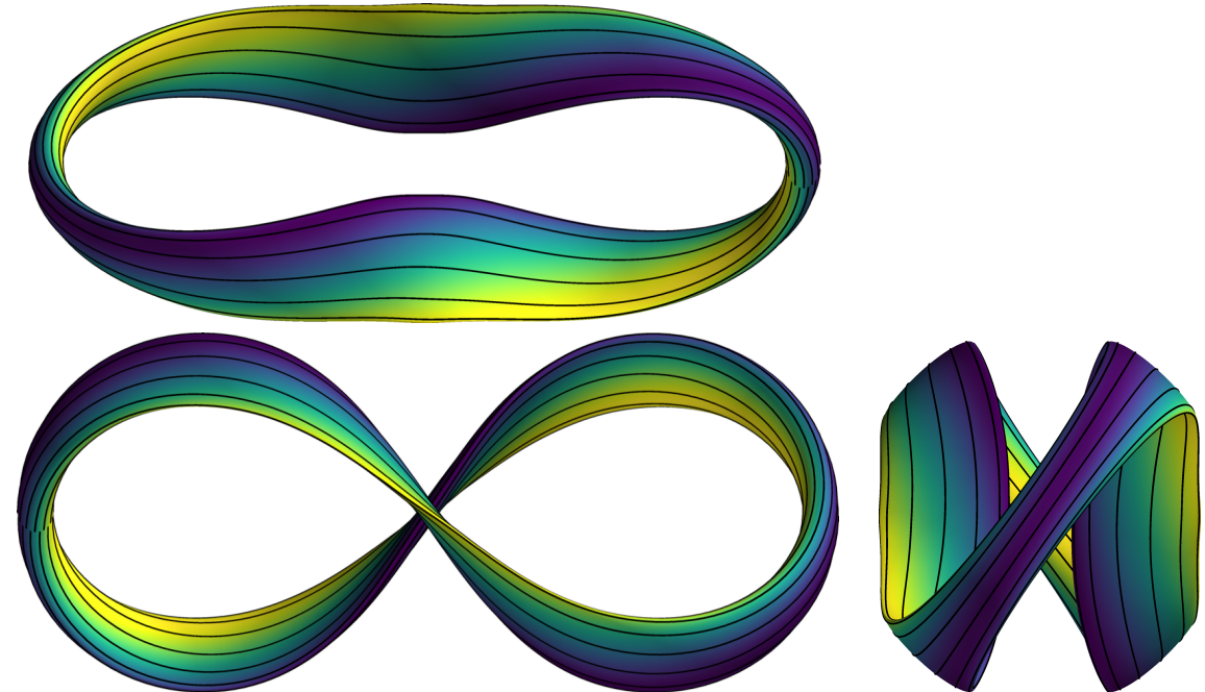} 
\caption{A quasi-helically symmetric stellarator surface with $|B|$ as obtained in \cite{landreman2022mapping}, seen from above (up-left), left side (down-left), and from the back (down-right). Reproduced from \cite{landreman2022mapping} with permission} \label{fig:StarngeShape1}
\end{minipage}\hspace{0.8cm}
\begin{minipage}{.4\textwidth}
\centering
\includegraphics[scale=0.3]{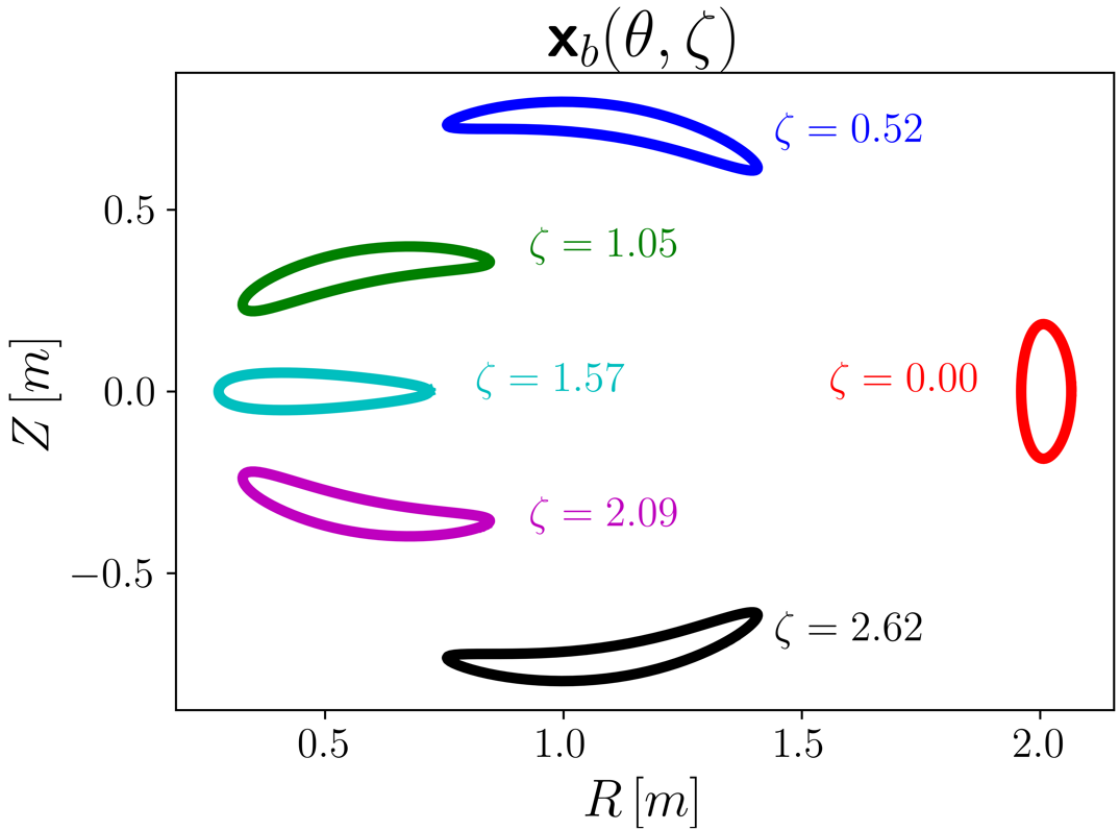}
\caption{Different cross sections for the same configuration in Figure \ref{fig:StarngeShape1} for different toroidal angles.}  \label{fig:StarngeShape2}
\end{minipage}
\end{figure}

Extreme 3D shaping pose a challenge for the numerical codes. Such boundaries are for example obtained from the near-axis expansion method \citep{jorge2020near}, for which an approximate semi-analytic expression for the equilibrium magnetic field is quickly obtained and the large
configuration space of stellarator equilibria can be rapidly explored and optimised \citep{rodriguez2023constructing}. Using the near-axis expansion, very attractive stellarator configurations have been discovered.
Examples include quasi-isodynamic configurations with small numbers of field periods \citep{landreman2021stellarator, jorge2022single, mata2022direct}, or the first reported quasi-helically symmetric configuration to have only 2 field-periods  \citep{landreman2022mapping} (see Figures \ref{fig:StarngeShape1} and \ref{fig:StarngeShape2}).

These highly optimised configurations are invariably also strongly shaped. They are often characterised by strong poloidal elongations with significant ellipticity and torsion from the magnetic axis. The more exact MHD equilibrium codes are still crucial for globally validating the near-axis results, posing a challenge; indeed, for the case shown in Figures \ref{fig:StarngeShape1} and \ref{fig:StarngeShape2}, some of the codes have been reported to fail producing overlapping flux surfaces.

The problem was identified to
be associated with how each code initialises the equilibrium calculation by interpolating the boundary from the axis, leading to overlapping coordinate surfaces. An improved method for choosing the coordinate axis proposed in \citet{qu2020coordinate} partially prevented coordinate surfaces from intersecting. However, this procedure still fails for arbitrarily strongly shaped boundaries. In more mathematical terms, we seek the conformal map of a simply connected domain, with a Jordan curve as a boundary, to the unit disc. The mapping is proven to exist (the Osgood–Carathéodory theorem \citep{henrici1993applied}); however, finding the numerical conformal map is not trivial, as shown in \citet{trefethen2020numerical} and \citet{porter2005}.

In this paper, we propose a new method for constructing polar-like and boundary-conforming coordinates, which enables the computation of grids in smooth and non-intersecting strongly shaped boundaries by exploiting a variational principle. Despite providing a solution to a plasma-physics-driven problem, this construction can be applied to other fields where coordinate mappings in strongly shaped cross sections are needed, e.g. for blood flow simulation \citep{zhang2007patient}.

The paper is organised as follows. In section \ref{Sec:continuousActionIntegral}, we introduce an action $\mathcal{S}$ and we develop the formalism for the mapping given as input the boundary surface, illustrating the main reasons why we expect the extremal points of the functional to be consistent with a mapping for which the coordinate surfaces are non-overlapping. 
In section \ref{Sec:ELderivation}, looking at the stationary points in $\mathcal{S}$, we derive and comment the Euler-Lagrange (EL) equations. 
In section \ref{Sec:numericalimplementation}, representing the mapping in a Fourier-Zernike basis and exploiting the geometrical meaning of the core terms in the Lagrangian, we provide a numerical method for the coordinate construction. In section \ref{Sec:Results}, we illustrate the results obtained by starting from ill-posed coordinate grids both with axisymmetric and non-axisymmetric external boundaries. 
The application of our coordinate construction to the computation of vacuum 3D equilibria by GVEC is illustrated in section \ref{Sec:GVECapplication}. Finally, a discussion is presented in section \ref{Sec:Conslusions}.

 

%% file: Sections/Mathematical_Formalism.tex
\section{Continuous action integral} \label{Sec:continuousActionIntegral}

We will consider general mappings $\textbf{x}(s, \theta, \zeta)$ between $(s, \theta, \zeta)$ with $s \in [0, 1]$, $\theta$, $\zeta\, \in \, [0, 2 \pi)$, and $\mathbb{R}^3$, analytic, and double periodic in $\theta$, and $\zeta$, $\textbf{x}(s, \theta, \zeta) = \textbf{x}(s, \theta + 2\pi, \zeta) = \textbf{x}(s, \theta, \zeta + 2 \pi)$. 
Boundary conditions are realised by providing a smooth toroidal external boundary $\textbf{x}_b(\theta, \zeta)$ and by requiring that $\textbf{x}(1, \theta, \zeta) = \textbf{x}_b(\theta, \zeta)$. 
A coordinate axis is defined as $\textbf{x}_a(\zeta) = \textbf{x}(0, \theta, \zeta)$, demanding that the mapping continuously changes its topological structure from surfaces with $s \neq 0$ to a curve in space at $s = 0$. 
This condition creates a coordinate singularity at $\textbf{x}_a$, since for fixed $\zeta$ every $\theta$ corresponds to the same point in real space, analogously to the polar coordinates. 
The objective of this work is to provide a construction of $\textbf{x}$ such that it becomes a \emph{polar-like coordinate mapping}, defined by having a non-vanishing coordinate Jacobian outside of the coordinate axis. Given such mapping 
$\textbf{x}(s, \theta, \zeta)$, we can compute the covariant basis vectors $\textbf{e}_s = \partial_s \textbf{x},\, \textbf{e}_\theta = \partial_\theta \textbf{x}, \, \textbf{e}_\zeta = \partial_\zeta \textbf{x}$, along with the corresponding metric elements $g_{ij} = \textbf{e}_{i}\cdot\textbf{e}_j$ with $i,j\, = s, \theta, \zeta$. 
The Jacobian is given by $\sqrt{g} \equiv \textbf{e}_s \cdot (\textbf{e}_\theta \cross \textbf{e}_\zeta)$. The contravariant elements are $\nabla s = \textbf{e}_\theta \times \textbf{e}_{\zeta}/\sqrt{g}$, $\nabla \theta = \textbf{e}_\zeta \times \textbf{e}_{s}/\sqrt{g}$, and $\nabla \zeta = \textbf{e}_s \times \textbf{e}_{\theta}/\sqrt{g}$. Considering different explicit forms of $\textbf{x}$ in the space of mappings conforming to a boundary leads to different forms of the metric elements.

We now introduce a \emph{geometrical action} $\mathcal{S}$ for the mapping $\textbf{x}$ constrained to the boundary $\textbf{x}_b$ as:
\begin{equation}\label{eq:Action}
     \mathcal{S}[\textbf{x}]  = \integral\left( \frac{1}{2} f \sqrt{g}^2 + \omega \sqrt{\textbf{e}_s \cdot \textbf{e}_s}\right).
 \end{equation}
$f(\textbf{x}; s, \theta, \zeta)$ is a positive function of the mapping and it can \emph{explicitly} depend on $(s, \theta, \zeta)$.  As discussed in Section \ref{Sec:ELderivation}, $f$ can be used for fixing the functional form of $\sqrt{g}$. A specific form for $f$ given below is chosen to recover the simple case of circular cross section axisymmetric torus to ensure the polar-like property of $\textbf{x}$. $\omega$ is a constant weighting coefficient. We assert that the minimum of $\mathcal{S}$ in the space of mappings maximises the set of points in coordinate space where the Jacobian is non-zero.

The corresponding Lagrangian $\mathcal{L} = \frac{1}{2} f \sqrt{g}^2 + \omega \sqrt{\textbf{e}_s \cdot \textbf{e}_s}$ encodes the key geometrical insights that lead to the desired coordinate mapping property. From the analyticity of $\textbf{x}$, for having $\sqrt{g} \neq  0$, it is crucial for the Jacobian to have a uniform sign. 
Considering that the overall volume $\mathcal{V} \int d\zeta d\theta ds \,\sqrt{g} $ is independent of the mapping, if $\sqrt{g}$ was decreasing up to becoming negative in a given subset of volume, it must increase somewhere else to counteract the effect, therefore, leaving $\mathcal{V}$ constant. The minimisation of $\sqrt{g}^2$ smooths the regions where $\sqrt{g}$ has a peak in its amplitude. Due to volume conservation, the gradients near the regions where the Jacobian is changing sign get weakened up to the point where it reaches homogeneity in sign. The remaining proportionality to $\sqrt{g}^2$, represented by $f$, needs to have a constant sign. Adding  $ \omega \sqrt{\textbf{e}_s \cdot \textbf{e}_s}$ to the action serves to penalise curvature, and thus increasing $\omega$ leads to coordinate mappings with straighter coordinate lines. This contribution to the action was found to be required after an initial numerical exploration suggested that, otherwise, the action functional would be independent of the parametrisation in the poloidal angle coordinate.

Note that there are multiple ways to include the boundary constraint $\textbf{x}(1, \theta, \zeta) = \textbf{x}_{b}(\theta, \zeta)$, e.g. by introducing a Lagrange multiplier $\lambda(\theta, \zeta)$ in $\mathcal{S}$ via the addition of the term
 \begin{align}
     \int_0^{2\pi}\!\!\!\!\! d\zeta \int_{0}^{2\pi}\!\!\!\!\! d\theta\, \lambda(\theta, \zeta)|\textbf{x}(1, \theta, \zeta) - \textbf{x}_b(\theta, \zeta)|.
 \end{align}
 
\section{Euler-Lagrange equations} \label{Sec:ELderivation}
In the variational approach,
$\textbf{x}$ constitutes the dynamic variable independent from $(s, \theta, \zeta)$. The relation between $\textbf{x}$ and $(s, \theta, \zeta)$, i.e. the mapping,  is recovered by solving the Euler-Lagrange (EL) equations, analogously to the Lagrangian approach in classical mechanics.
We can compute the stationary points in the action $\mathcal{S}$ in Eq. (\ref{eq:Action}) by considering the variation of the coordinate mapping $\delta \textbf{x} \equiv \textbf{x} - \textbf{x}^{'}$, where $\textbf{x}$ and $\textbf{x}{'}$ are two independent mappings constrained to $\textbf{x}_b$. 
The boundary conditions fix the possible mapping variations at the external surface, $\delta \textbf{x}(1, \theta, \zeta) = \delta \textbf{x}_b(\theta, \zeta) = 0$; and at the coordinate axis, $\delta \mathbf{x}(0, \theta, \zeta)$ = $\delta \mathbf{x}(\zeta)$.
The variation of the action in Equation (\ref{eq:Action}) with respect to the coordinate mapping $\textbf{x}(s, \theta, \zeta)$ yields to the EL equations:
\begin{equation}\label{ELfull}
\nabla (f \sqrt{g}) - \frac{\sqrt{g}}{2} \frac{\delta f}{\delta \textbf{x}} + \frac{\omega}{\sqrt{g}}\boldsymbol{\kappa}= 0,
\end{equation}
having defined the curvature vector as $\boldsymbol{\kappa} = (\textbf{e}_s \cdot \nabla) \textbf{e}_s$. It accounts for the variation of the tangential lines along the radial direction, being a measure of the straightness of $\theta,\zeta=$constant coordinates lines. The EL equations are a set of nonlinear partial differential equations with boundary conditions. Their derivation is presented in Appendix \ref{Appendix1}.
The first term on the left of Eq. (\ref{ELfull}) accounts for the volume elements compared to a Jacobian that would scale as $1/f$. The second term is the contribution from the variation of $f$ with respect to $\textbf{x}$. Lastly, the curvature term represented by 
$\boldsymbol{\kappa}$ and weighted by $\omega$, quantifies deviations from straight radial lines. 
  
The function $f$ can be either externally prescribed or determined by solving Equation (\ref{ELfull}) for a specific choice of $\textbf{x}$. To fix $f$, we impose that the known analytical mapping of a torus with a circular cross-section is recovered exactly by the action minimisation. The analytical mapping reads as $\textbf{x} = R(\cos \phi \textbf{e}_x + \sin \phi \textbf{e}_y) + Z \textbf{e}_z$, with $\textbf{e}_x, \textbf{e}_y, \textbf{e}_z$ unitary Cartesian vectors, and $R = 1 + s \cos(\theta),Z=s\sin(\theta)$, with $\theta$ representing the geometrical poloidal angle. 
By choosing 
\begin{equation}
    f = 1/s(R(\textbf{x}, s, \theta, \zeta))^2\,, \label{eq:fchoice}
\end{equation}
we have $\delta f/ \delta \textbf{x} = - \frac{2}{s R^3} \hat{\textbf{R}}$. Substituting this expression into Eq. (\ref{ELfull}), the Euler-Lagrange equations are trivially satisfied, as intended. Given the topological equivalence of toroidal domains with a torus with a circular cross-section, we choose   \eqref{eq:fchoice} to define $f$ for the following analysis.


%% file: Sections/Numerical_Implementation.tex
\section{Numerical Implementation} \label{Sec:numericalimplementation}

\subsection{Zernike-Fourier representation}

We consider the local form for the mapping as $\textbf{x} = R(\textbf{x}, s, \theta, \zeta)\,\hat{\textbf{R}} + Z(\textbf{x}, s, \theta, \zeta)\, \textbf{e}_z$. 
The jacobian is computed as: $\sqrt{g} = R \sqrt{g}_p =  R \big(\partial_s R \partial_\theta Z -  \partial_s Z \partial_\theta R\big)$, where $\sqrt{g}_p$ is the poloidal part of the jacobian. The metric elements are $g_{ij} = \partial_i R \partial_j Z + \partial_j R \partial_i Z + \delta_{i\zeta} R^2$, where $\delta_{i\zeta} = 1$ if $i = \zeta$ and $0$ otherwise. In addition to considering double periodic mappings, we impose stellarator symmetry \citep{lee1988optimum} for simplifying the computations. 
This discrete symmetry is defined as $R(s, -\theta, -\zeta) = R(s, \theta, \zeta)$ and $Z(s, -\theta, -\zeta) = -Z(s, \theta, \zeta)$, which implies a more general symmetry \citep{dewar1998stellarator}. Stellarators are typically composed of a number of identical sectors $N_{P}$, called field periods. 
A double-periodic function $h(s, \theta, \zeta)$ can be expressed in a Fourier-Zernike \citep{mcalinden2011mathematics} basis as:
\begin{align}
    h(s, \theta, \zeta) = \sum_{n = 0}^{+ \infty}\sum_{l = 0}^{+ \infty}\sum_{m = 0}^{l}\Bigg[&\Bigg( C_{+, nlm}\mathcal{Z}^{m}_l(s, \theta) + C_{-, nlm}\mathcal{Z}^{-m}_l(s, \theta) \Bigg)\cos(q N_{P} \zeta) \, + \\
    &  \Bigg(S_{+, qlm}\mathcal{Z}^{m}_l(s, \theta) + S_{-, nlm}\mathcal{Z}^{-m}_l(s, \theta) \Bigg)\sin(q N_{P} \zeta)\Bigg],
\end{align}
where the Zernike polynomials \citep{niu2022zernike} are given as $\mathcal{Z}^{m}_l = \mathcal{R}^m_l \cos(m \theta)$ and $\mathcal{Z}^{-m}_l = \mathcal{R}^m_l \sin(m \theta)$, with radial part

\begin{align}
  \mathcal{R}^m_l(s) = \begin{cases}
    \sum_{k = 0}^{\frac{l - m}{2}} \frac{(- 1)^k(l - k)!}{k!(\frac{l + m}{2}- k)!(\frac{l - m}{2} - k)!} s^{l - 2k}, & l - m \quad \text{even} \\
    0 , & l - m \quad \text{odd}
    \end{cases}
\end{align}
for $l\geq m$, with $n,l,m\, \in \, \mathbb{N}^{+}$. The real coefficients $C_{a, nlm} , C_{b, nlm}, S_{a, nlm}, S_{b, nlm}$ parametrise the mapping in the Fourier-Zernike space. The Zernike polynomials are orthogonal basis functions on the unit disk, and they can describe the mapping with half of the coefficients used by the parity-restricted Chebyshev polynomials \citep{mason2002chebyshev} often used in plasma physics. The normalisation is chosen such that the inner product $\langle A B \rangle = \int_0^1 \int_{0}^{2\pi} A(s, \theta) B(s, \theta) s ds d\theta$ defined on the space of $\mathcal{L}^2$ functions on the disk gives:  
\begin{align}
    \langle\mathcal{Z}^m_l \mathcal{Z}^{m'}_{l'} \rangle = \frac{1 + \delta_{m0}}{2(l + 1)} \delta_{m m'} \delta_{l l'}.
\end{align}
Stellarator symmetry imposes $\mathcal{C}_{b, nlm} = \mathcal{S}_{a, nlm} = 0$ for $R$ and $\mathcal{C}_{a, nlm} = \mathcal{S}_{b, nlm} = 0$ for $Z$. Extending the sum to negative values of $n$, and redefining the set of Fourier-Zernike coefficients in terms of $r_{qlm}, z_{qlm}$ and $q \in \mathbb{Z}$, the discretised mapping becomes, \citet{qu2020coordinate}:
\begin{align}
    & R(s, \theta, \zeta) = \sum_{n = -N}^{N}\sum_{m = 0}^{M}\sum_{m = l}^{L}  r_{nlm} \mathcal{R}^m_l(s)\cos(m \theta -  n N_{fp} \zeta), \label{eq:mappingR} \\
    & Z(s, \theta, \zeta) = \sum_{n = -N}^{N}\sum_{m = 0}^{M}\sum_{m = l}^{L} z_{nlm}  \mathcal{R}^m_l(s)\sin(m \theta -  n N_{fp} \zeta), \label{eq:mappingZ}
\end{align}
where $N$, $M$, and  $L$ $\in$ $\mathbf{N}^{+}$ indicate the resolution imposed in the Fourier-Zernike space. The number of components in $r_{nlm}$ and $z_{nlm}$ scales as $2 M^2 N$. The Fourier-Zernike mapping in the poloidal plane directly satisfies the condition to be analytic \citep{boyd2011comparing} since the radial part with poloidal mode number $m$ scales as $\mathcal{R}^m_l(s) \sim s^m(\mathcal{R}^{m}_{0} + \mathcal{R}^m_2 s^2 + \mathcal{R}^m_4 s^4 + \dots)$. Therefore, no special treatment in terms of coordinate regularisation is required for the coordinate axis, where the coordinate mapping will be polar-like at the leading order. The boundary surface $\textbf{x}_b(\theta, \zeta)$ is given in terms of its Fourier components,
\begin{align}
   &  R^b(\theta, \zeta) = \sum_{n = -N_b}^{N_b}\sum_{m = 0}^{M_b} R^b_{nm} \cos(m\theta - N_{P} n \zeta),\\ 
   & Z^b(\theta, \zeta) = \sum_{n = -N_b}^{N_b}\sum_{m = 0}^{M_b} Z^b_{nm} \sin(m\theta - N_{P} n \zeta),
\end{align}
where $N_b, M_b\, \in \, \mathbb{N}^{+}$ are the prescribed toroidal and poloidal boundary resolutions. In the following, we consider $M= L$. The boundary condition $\textbf{x}(1, \theta, \zeta) = \textbf{x}_{b}(\theta, \zeta)$ reduces the number of independent components in $r_{nlm}$ and $z_{nlm}$. Using the property for the Zernike radial part of $\mathcal{R}_l^m(1) = 1$ for each $m,l$, the explicit form for the boundary condition become:
\begin{align}
    \sum_{l = m}^{L}r_{nlm} = R^b_{nm}\, , \quad \sum_{l = m}^{L}z_{nlm} = Z^b_{nm}\,.
\end{align}
These equations are exploited to fix $(M + 1)(2N + 1)$ components of $r_{nlm}$ and $z_{nlm}$. 
The coefficients $r_{nmm}$ and $z_{nmm}$ for each $n$ and $m$ are indeed fixed as $r_{n mm} = R^{b}_{nm} - \sum_{l = m + 1}^{L}r_{nlm}$, $z_{n mm} = Z^{b}_{nm} - \sum_{l = m + 1}^{L}z_{nlm}$.

\subsection{Action discretisation}
\label{sec:action_discrete}

\begin{figure}
\centering 
\includegraphics[scale=0.4]{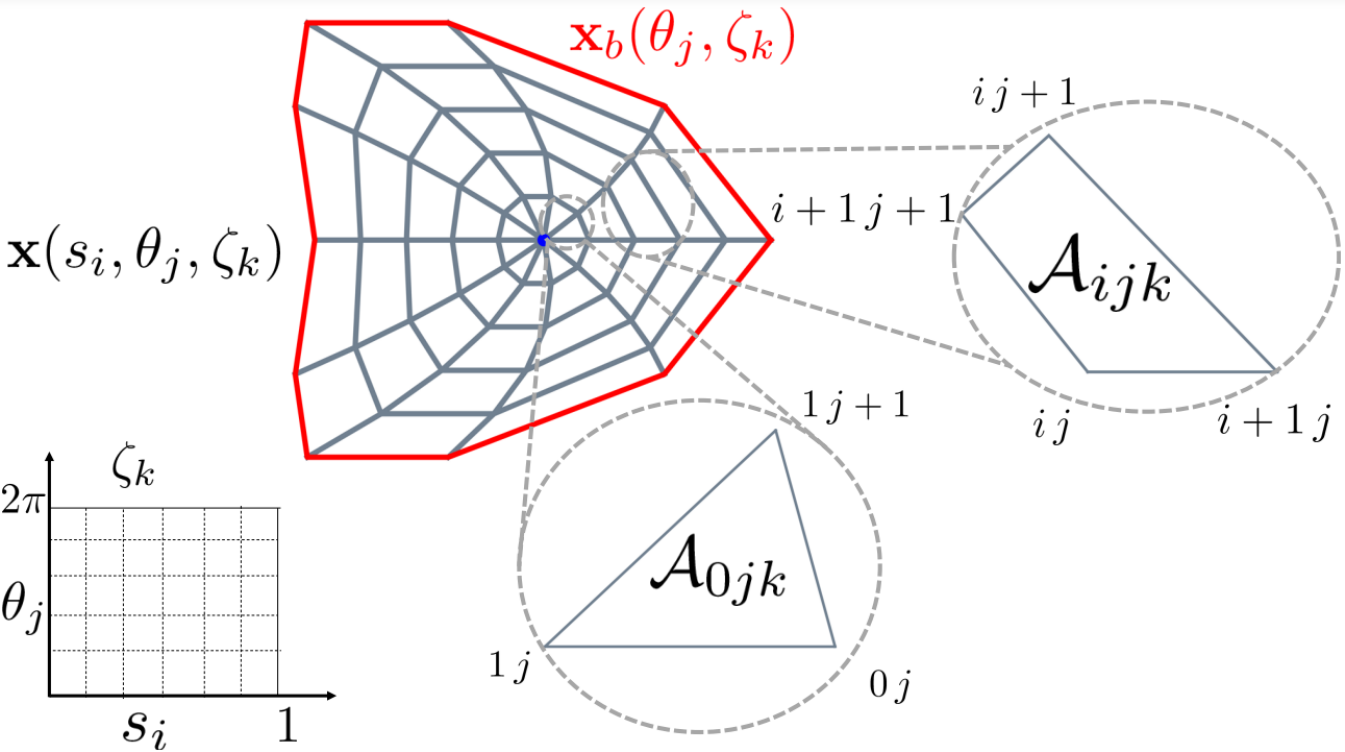}
\caption{Schematic illustration of the geometrical elements in the action discretisation}
\label{fig:Schetch}
\end{figure}

Instead of directly using the Zernike-Fourier representation of the mapping in Eq. (\ref{eq:Action}), exploiting a coordinate grid highlights the geometrical properties of $\mathcal{S}$ in terms of areas and radial lengths.
The idea is depicted in the sketch provided in Figure \ref{fig:Schetch}.

As drawn on the left, starting with a grid $\{s_i, \theta_j, \zeta_k\}$ in coordinate space with $i \in [0, N_s]$, $j \in [0, N_\theta]$, and $k \in [0, N_{\zeta}]$, fixing the value of $\zeta$, $\textbf{x}$ maps the corresponding points into a grid of $R_{ijk}$ and $Z_{ijk}$ in a poloidal plane. Connecting the points with straight lines along the coordinate lines we have a set of quadrilaterals and triangles. Each of them has an area $\mathcal{A}_{ijk}$ constituted by the grid points $\textbf{x}_{ijk}$, $\textbf{x}_{i + 1 j k}$, $\textbf{x}_{i + 1 j+1 k}$, and $\textbf{x}_{i j+1 k}$. At the coordinate axis, for which $i = 0$, we have that $\textbf{x}_{0 j+1 k} = \textbf{x}_{0 j k}$. The expression of $\mathcal{A}_{ijk}$ is explicitly given by the Shoelace formula \citep{braden1986surveyor}. Each point $\textbf{x}_{ijk}$ identifies a radial length $L_{ijk}$ to $\textbf{x}_{i+1jk}$ via the Euclidean distance of two points. The convention assumed for the identification of each grid point in $\mathcal{A}_{ijk}$ and in $L_{ijk}$ is shown in Figure \ref{fig:Schetch}. Given a point $\textbf{x}_{ijk}$ the points at $i + 1$ and $j + 1$ correspond to the incremental radial and poloidal coordinates, respectively. For this reason, $\mathcal{A}_{ijk}$ and $L_{ijk}$ are not defined at the boundary.

By choosing $f$ following Eq.\eqref{eq:fchoice}, the Jacobian reduces to a poloidal Jacobian, such that $f \sqrt{g}^2 = \sqrt{g}_p^2/s$. Thus, the \emph{discrete} action integral can be computed as:
\begin{align}\label{eq:Sdiscretised}
    \mathcal{S}[\textbf{x}_{ijk}] = \sum_{ijk}\Big(\frac{1}{2 s_i}  \mathcal{A}^2_{ijk} +  \omega L_{ijk}\Big),
\end{align}
where we approximated the poloidal Jacobian with the area element, considered an equispaced grid, and neglected common multiplicative factors. $\mathcal{S}[\textbf{x}_{ijk}]$ is an explicit function in terms of the free variables $\{r_{nlm}, z_{nlm}\}$. Minimising $\mathcal{S}$ is an unconstrained nonlinear optimisation problem that we solve using the Broyden-Fletcher-Goldfarb-Shanno algorithm \citep{dai2002convergence}, a descent-gradient iterative method with an explicit computation of the gradient in the space of free parameters as illustrated in Appendix \ref{sec:AppendixD}.

%% file: Sections/Results.tex
\section{Results} \label{Sec:Results}

\subsection{Axisymmetry}

\begin{figure}
 \centering
\includegraphics[width=0.6\textwidth]{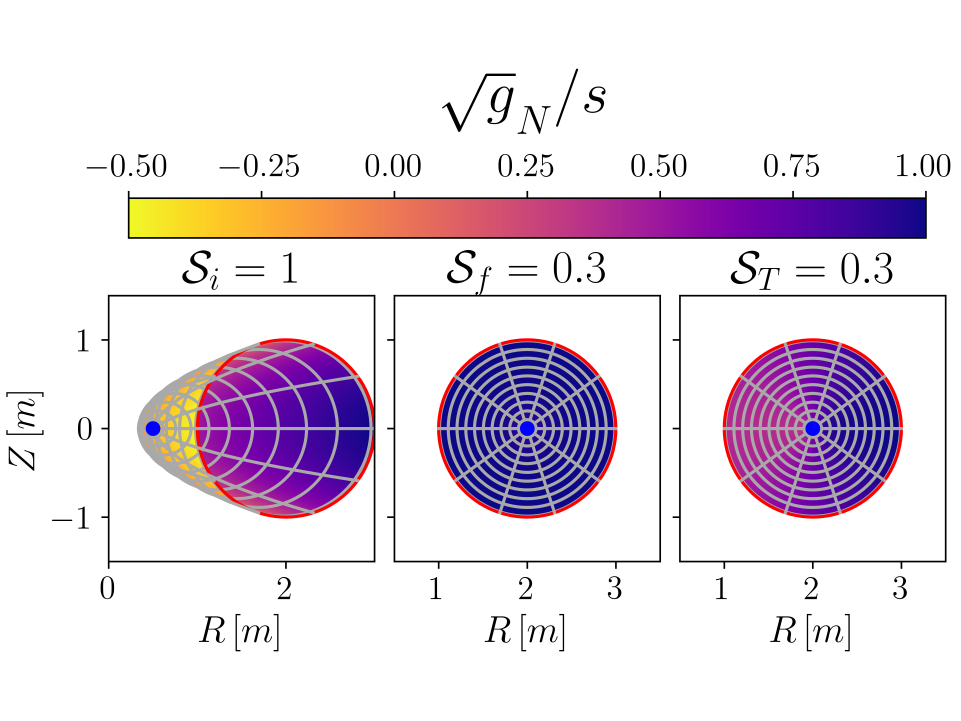}
 \caption{Normalised Jacobian $\sqrt{g}_N/s$ (in colour) and constant coordinate lines  (in grey) for a circular torus and the external boundary (in red). The coordinate axis is dotted in blue. The initial grid on the left with initial action value $\mathcal{S}_i$ is compared to the optimisation outcome with $\mathcal{S}_f$. In the right plot, the 3D Jacobian is shown (including $R$). }
 \label{fig:3circles}
 \end{figure}
 
To validate the action-based approach, we first consider the coordinate construction for axisymmetric boundary surfaces both analytically and numerically. Exploiting the inherent symmetry along the toroidal angle, we transform the problem into one that is focused on optimising the jacobian within a single poloidal plane. The toroidal index is left out for simplicity.

We consider the family of boundaries, proposed in \citet{qu2020coordinate}, parameterised as $R^b(\theta) = R_0^b + R_1^b \cos \theta + R_2^b \cos 2\theta$ and $Z^b(\theta) = Z_0^b + Z_1^b \sin \theta + Z_2^b \sin 2\theta$. We consider a poloidal mapping $\textbf{x}(s, \theta)$ with a Zernike polynomial of order $L = 2$. Combining with boundary conditions, we get: $R(s, \theta) = R_0^b + 2 r_{20}(s^2 - 1) + R_1^b  s \cos \theta + R_2^b s^2 \cos 2 \theta$, $Z(s, \theta) =  Z_0^b + Z_1^b  s \sin \theta + Z_2^b s^2 \sin 2 \theta $. The only degree of freedom is  $r_{20}$.
The action $\mathcal{S}[r_{20}]$ is a second-order polynomial in $r_{20}$ plus a contribution from the integrated radial length proportional to $\omega$. In the numerical implementation we set $M = L = 2$, $N_{s} \times N_\theta = 1000 \times 1000$.

Firstly, as a sanity check for the numerical implementation, we take $\textbf{x}_b$ as a torus with circular cross-section, $R_0^b = 2$, $R_1^b = Z_1^b = 1$ and $R_2^b = Z_2^b = 0$. The action becomes $\mathcal{S}[r_{20}] = \pi( 1 + 4 r_{20}^2) + \omega \int_{0}^{2\pi}d\theta \int_0^{1} ds \sqrt{16 r_{20}^2 s^2 + 8 s r_{20} \cos \theta + 1}$. 
The study of $\mathcal{S}$ shows the presence of a global minimum for $r_{20}$ between $0$ and $10^{-10}$ for $\omega$ varying between $0$ and $10$. Figure \ref{fig:3circles} shows the starting and optimised grids, with a normalised Jacobian $\sqrt{g}_N/s=\sqrt{g}/(\max(\sqrt{g})s)$, being unitary for the circular polar grid. The starting grid on the left has been chosen with an initial coordinate axis outside the external boundary, creating a region with ill-posed coordinates, as $\sqrt{g}_N$ becomes negative with overlapping coordinate lines. As shown in Figure \ref{fig:3circles}, the algorithm finds the polar coordinate in the poloidal plane as expected, with $\sqrt{g}_N/s$ being unitary in all the domain. $\textbf{x}(s, \theta, \zeta)$ retrieves the toroidal coordinate with Jacobian $s (R_2^b + s \cos \theta)$ as in Figure \ref{fig:3circles} on the right. The plotted results are for $\omega = 10^{-1}$, and its variation does not affect the final result significantly as the analytical discussion anticipated. 

\begin{figure}
\begin{minipage}[t]{0.48\textwidth}
\includegraphics[valign=t, width=1\linewidth]{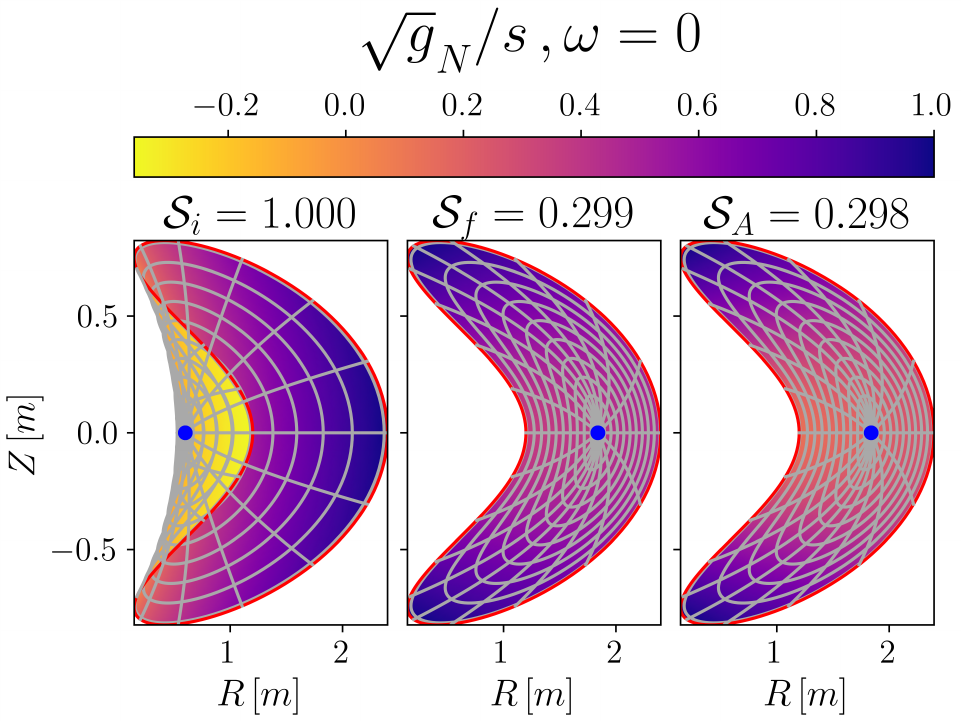} 
\caption{Normalised Jacobian $\sqrt{g}_N/s$ (in colour) and constant coordinate lines (in grey) for the bean-shaped external boundary (in red). The coordinate axis is dotted in blue. The initial grid on the left with initial action value $\mathcal{S}_i$ is compared to the outcome of the optimisation with $\mathcal{S}_f$ and the analytical result $\mathcal{S}_A$.}  \label{fig:AnaliticComp}
\end{minipage}\hspace{0.5cm}
\begin{minipage}[t]{0.48\textwidth}
\includegraphics[valign=t, width=1\linewidth]{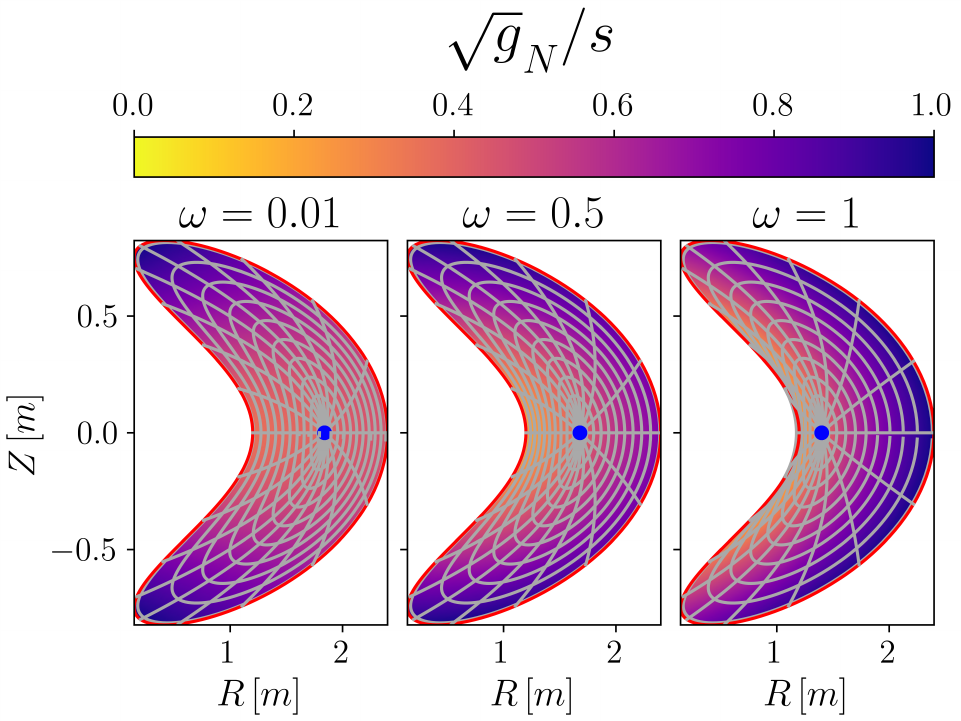} 
\caption{Comparison of optimised configurations for increasing values of $\omega$. Plotted are the normalised Jacobian $\sqrt{g}_N/s$ (in colour) and constant coordinate lines (in grey) for the bean-shaped external boundary (in red). The coordinate axis is dotted in blue.}  \label{fig:Straight}
\end{minipage}
\end{figure}
 
Next, the strongly shaped boundary is considered a bean-shaped contour described by $R_0^b = 1 $ $Z_b^0 = 0$, $R_b^1 = 0.6$, $Z_b^1 = 0.8$, $R_b^2 = 0.8$ and $Z_b^2 = 0.1$. Considering $\omega = 0$, the minimum of the poloidal action is found analytically at $r_{20} = -0.42$. Initiated with $r_{20} = 0.2$, the coordinate lines in the initial grid, shown in Figure~\ref{fig:AnaliticComp}, exhibit significant overlap, resulting in a negative Jacobian. The minimisation algorithm yields a valid mapping aligned with the analytical prediction, as the centre and right plots in Figure ~\ref{fig:AnaliticComp} illustrate. Subsequent increments in the $(s, \theta)$ grid resolution leads to the minimisation of the action $\mathcal{S}_{f}$ asymptotically reaching the analytical value $\mathcal{S}_A = 0.298$. Low values for the radial weight $\omega$ allow for more curved coordinate lines, as Figure~\ref{fig:Straight} shows, whereas increasing $\omega$ straightens radial coordinate lines and shifts the coordinate axis to the left.

\subsection{Non-axisymmetry}

\begin{figure}
 \centering
\includegraphics[width=0.75\textwidth]{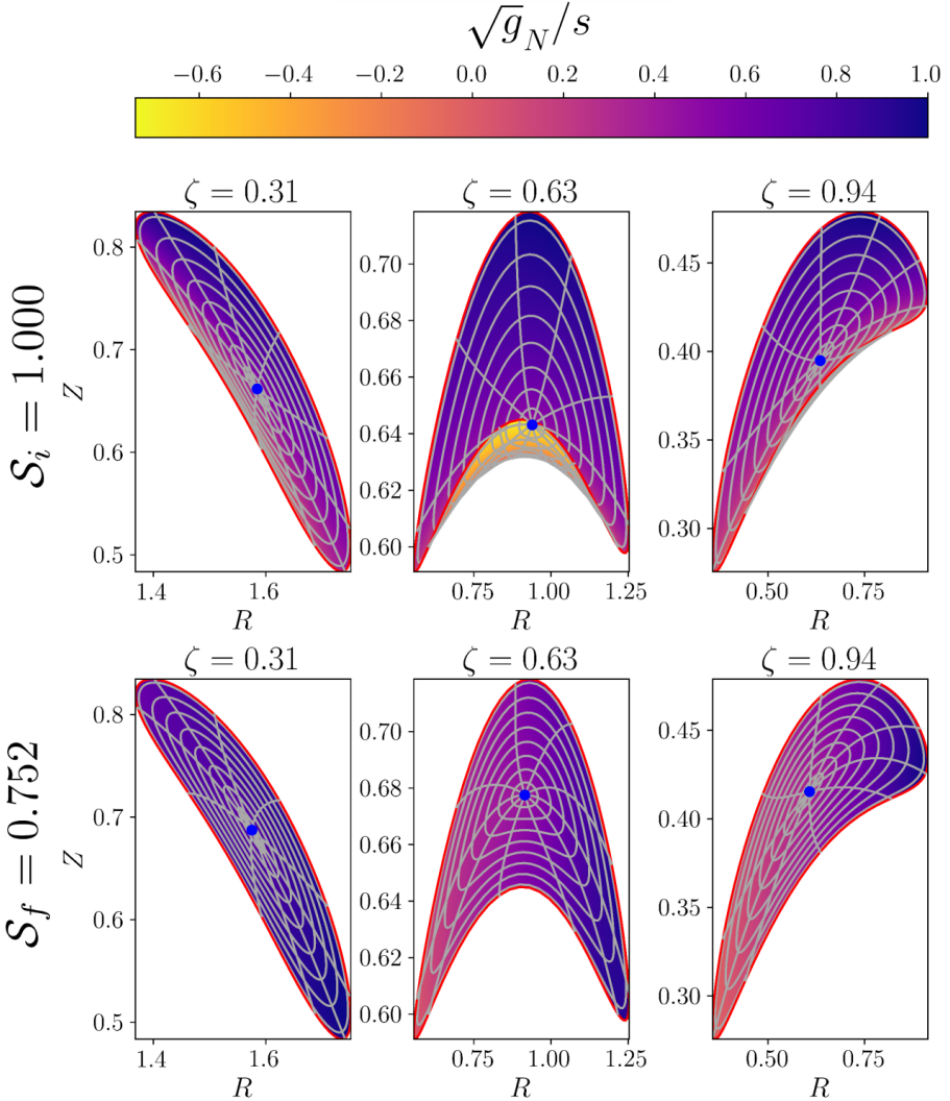}
 \caption{Comparison between the starting (top) and optimised (bottom) configurations at different toroidal planes. Plotted are the normalised Jacobian $\sqrt{g}_N/s$ (in colour) and constant coordinate lines (in grey) for a strongly shaped boundary (in red). The coordinate axis is dotted in blue.}
 \label{fig:3Dplot}
 \end{figure}

The strongly shaped boundary in Figure \ref{fig:StarngeShape1} provides a non-axisymmetric test where the coordinate construction using a polynomial interpolation from the boundary fails. The optimisation here is performed with $M = N = L = 5$ and $N_{s} \times N_\theta \times N_\zeta =  80 \times50 \times 80$ with $\omega = 10^{-2}$. The initial grid is computed by setting all the degrees of freedom to zero. Figure \ref{fig:3Dplot} shows that coordinate singularities in the initial grids (top panel with starting action value $\mathcal{S}_i = 1$) are avoided by the optimisation algorithm (bottom panel with final action value $\mathcal{S}_f = 0.752$). We would like to remark that, given a value of $\omega$ and a grid resolution, the algorithm finds this same optimal mapping no matter how bad the initial guess is (data not shown). The same happens in the axisymmetric case, showing the algorithm's robustness. 

\subsection{Numerical verification of EL equations}

Given a mapping $\textbf{x}$ we can provide a measure of its distance from the theoretical extremal point of $\mathcal{S}$ via the EL equations (\ref{ELfull}). Defining the operator $\boldsymbol{\mathcal{L}}(\textbf{x})$ as the left side of Eq. (\ref{ELfull}), we consider the volume average, 
\begin{align}
    \langle s^4 \boldsymbol{\mathcal{L}}^2 \rangle = \frac{1}{V}\integral \sqrt{g} s^4 \boldsymbol{\mathcal{L}}^2,
\end{align}
where $V$ is the volume which depends solely on the boundary. $\langle s^4 \boldsymbol{\mathcal{L}}^2 \rangle$ globally quantifies the deviation of the mapping from the extremal solution. Introducing the $s^2$ factor ensures numerical stability near the axis, as $\dv{f}{s} \sim 1/s^2$.

\begin{figure}
\begin{minipage}[t]{0.49\textwidth}
\centering
\includegraphics[valign=t, width=1\linewidth]{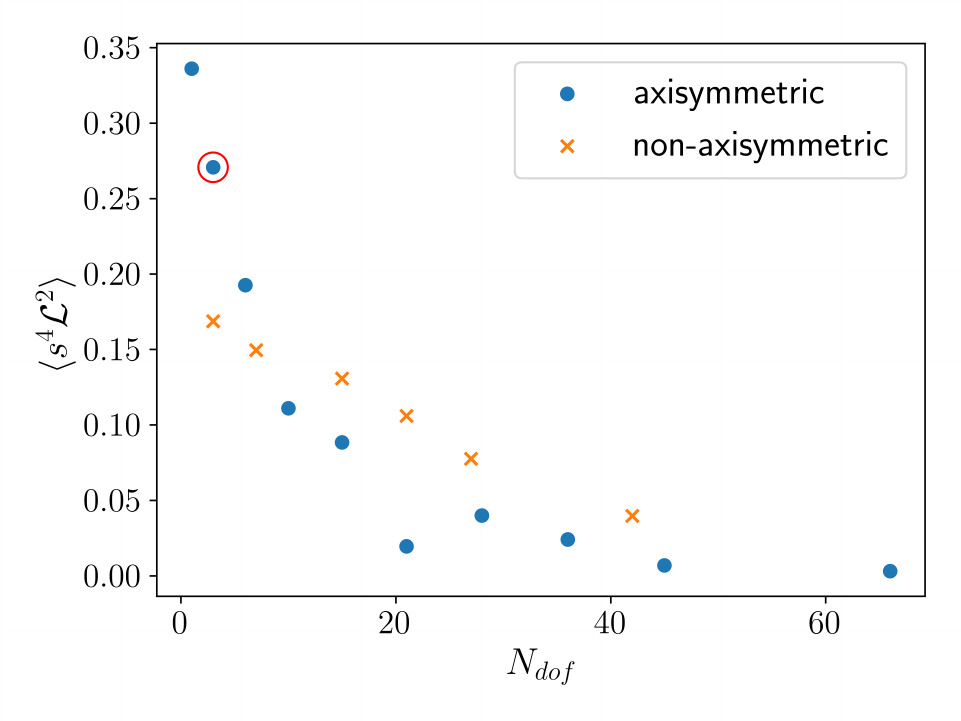} 
\caption{convergence of the solution from the discretisation of the numerical action with the increase in the free parameters to the EL stationary point for an axisymmetric case and a non-axisymmetric one.}\label{fig:ELleft}
\end{minipage}
\hfill
\begin{minipage}[t]{0.49\textwidth}
\centering
\includegraphics[valign=t, width=1\linewidth]{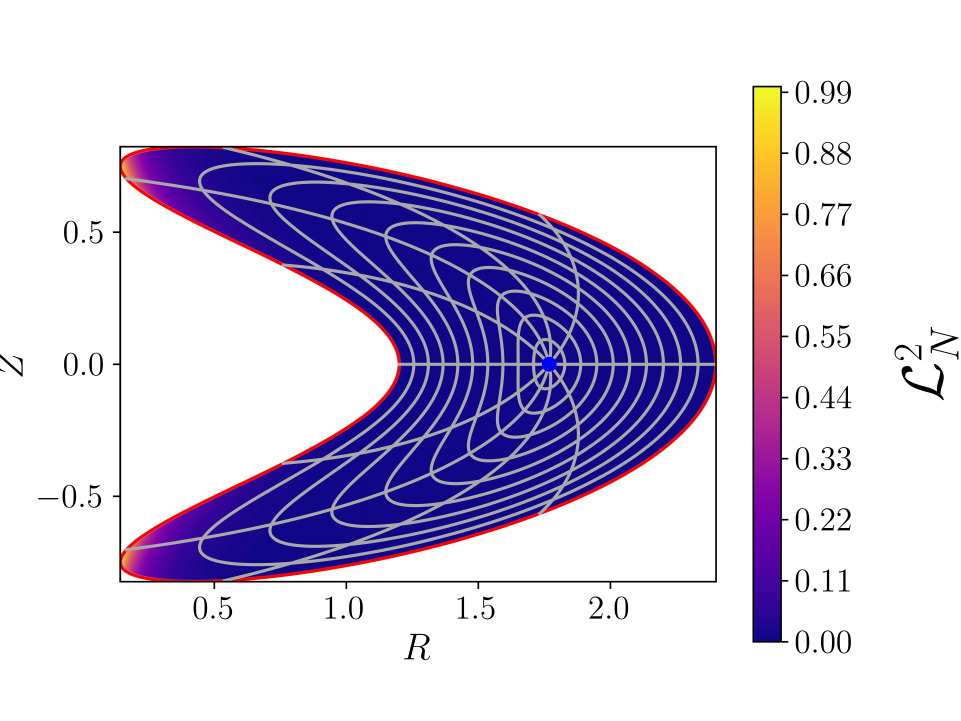} 
\caption{$\mathcal{L}^2$ is analysed locally for $N_{dof} = 3$ (red circle of Figure ). The coloured regions show $\mathcal{L}^2$ normalised to its maximum value.}\label{fig:ELright}
\end{minipage}
\end{figure}

In Figure \ref{fig:ELleft}, we observe that increasing the Zernike-Fourier discretisation, equivalent to having a higher number of free parameters $N_{dof}$ in the optimisation, results in a better approximation of the mapping solution, both in axisymmetric and non-axisymmetric cases for strongly shaped boundaries. Notably, $\mathcal{L}^2$ deviates from zero primarily near the boundary and where the coordinate lines exhibit rapid variation compared to inner regions (Figure \ref{fig:ELright}). The boundary coefficients used for the numerical convergence study in Figure \ref{fig:ELleft} can be found in Appendix \ref{ultima}. 

%% file: Sections/GVEC_application.tex
\section{GVEC application} \label{Sec:GVECapplication}
\begin{figure}
 \centering \includegraphics[width=0.95\textwidth]{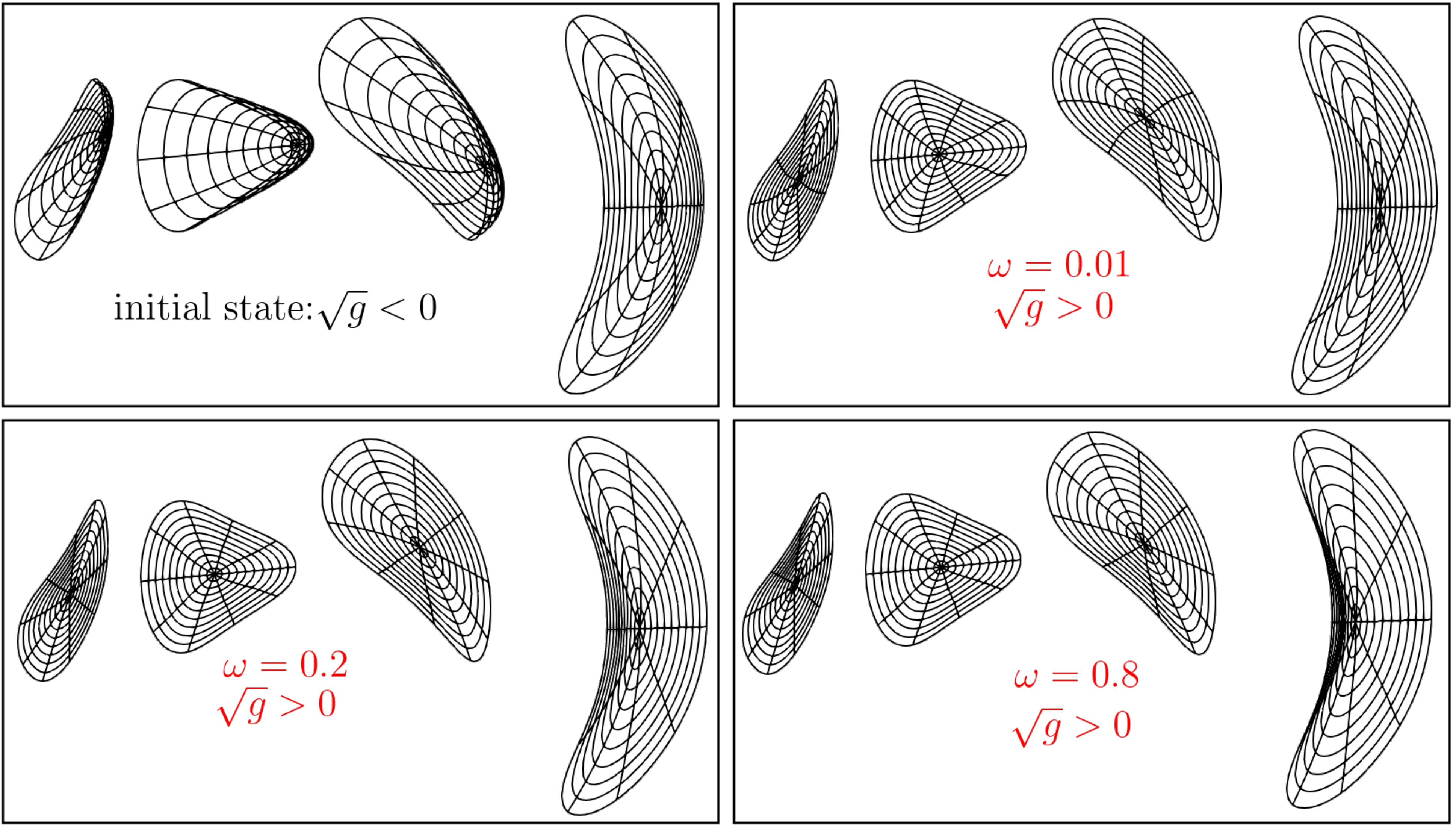}
 \caption{GVEC results for the minimisation of the {\it action functional} \eqref{EL} using a 3D W7-X like boundary and a circular axis for initialisation. The $(s,\theta)$ grid is shown for 4 cross-sections, and the mapping is discredised with a Zernike - Fourier representation, using $1$ B-Spline element in the radial direction and $(m,n)_\textrm{max}=7$. Top left: Initial state with $\sqrt{g}<0$. Top right and bottom: valid optimised grids for increasing weighting factor $\omega$. }
 \label{fig:GVEC_w7x_case}
 \end{figure}

The approach of minimising the action functional \eqref{EL} is successfully tested in GVEC \citep{gvec_simons}, a 3D MHD equilibrium solver assuming closed nested flux surfaces. In GVEC, the flux surface coordinate mapping $R(s,\theta,\zeta), Z(s,\theta,\zeta)$ is discretised by a tensor-product of B-splines in $s$ and Fourier-modes in angular directions. Using a $1$ element B-Spline of degree equal to the maximum poloidal mode number, and smoothness constraints at the axis, the Zernike-Fourier representation is exactly recovered. As a 3D test case, a W7-X-like boundary with an initial circular magnetic axis was taken, where the usual initialisation of GVEC produces $\sqrt{g}<0$ (Figure \ref{fig:GVEC_w7x_case}). 

The minimisation of $\mathcal{S}$ with a gradient-descent successfully produced $\sqrt{g}>0$.
This is shown in Figure~\ref{fig:GVEC_w7x_case}, where the initial invalid $(s,\theta)$ grid and final valid grids for different weighting factor $\omega$ are shown. There is a trade-off between the straightness of the $\theta$ contours and the axis position, especially in the bean-shaped cross-section. Using these solutions as initialisation for the MHD energy minimisation, it was found that GVEC had no difficulties converging to an equilibrium for moderate $\omega\leq0.2$. 


%% file: Sections/Conclusions.tex



 \section{Discussion}\label{Sec:Conslusions}

This paper introduces a method for constructing polar-like and boundary-conforming coordinates within arbitrarily shaped toroidal domains based on identifying a mapping that extremises an action formulated from geometric principles. Preserving bijectivity follows from minimising the squared jacobian (scaled by a positive factor $f$) and the radial curvature of the mapping. The first variation of $\mathcal{S}$ yields a set of nonlinear partial differential equations, balancing a gradient comparing the relative scaling of $f$ and $\sqrt{g}$, a radial curvature term, and the variation of $f$ with respect to the mapping. $f$ is determined using a topological argument with a torus with a circular cross-section in cylindrical coordinates as a reference.

Both axisymmetric and non-axisymmetric tests, conducted through analytical and numerical means, demonstrate the effectiveness of minimising $\mathcal{S}$ in constructing coordinates for domains with complex shapes. As the poloidal and toroidal discretisation increases, the volume-average $\langle s^4 \boldsymbol{\mathcal{L}}^2 \rangle$ indicates convergence of the optimised mapping towards the Euler-Lagrange zero, with notable discrepancies observed primarily in regions near the boundary where significant variations occur and high resolution is essential. The robustness of the action formalism is further validated by its application in computing 3D MHD equilibria within the GVEC framework \citep{gvec_simons}, emphasising its efficacy where alternative coordinate choices need a careful a-priori choice of the initial axis position, without the guarantee to yield a valid mapping.

A thorough study of the existence of global minima for the action in Equation (\ref{eq:Action}), along with an exploration of solutions to the EL equations \ref{ELfull}, could significantly expedite the computation of the optimal coordinate mapping. Ongoing analysis is directed towards studying the action formalism in the formalism of harmonics maps \citep{eells-sampson-1964,hamilton-1975} with their close relation to conformal mappings. Analysing the influence of parameters $f$ and $\omega$ on these solutions can provide insights for enhancing the jacobian in specific regions of a given boundary. While directly solving the EL equations (\ref{ELfull}) instead of employing a global approach based on minimising an action integral may offer faster convergence, the inherent challenge persists due to the nonlinear nature of the problem and the prescribed strongly shaped boundary conditions.

The findings presented in this paper hold potential value for 3D MHD equilibrium codes and other codes utilising a global set of toroidal flux coordinates. This method can facilitate the exploration of complex, strongly shaped configurations that were previously challenging to address or enhance the convergence and precision of existing geometries across various applications.

%% file: Sections/Acknowledgments.tex
\section*{Acknowledgements}
This work has been carried out within the framework of the EUROfusion Consortium, via the Euratom Research and Training Programme (Grant Agreement No 101052200 — EUROfusion) and funded by the Swiss State Secretariat for Education, Research and Innovation (SERI). Views and opinions expressed are however those of the author(s) only and do not necessarily reflect those of the European Union, the European Commission, or SERI. Neither the European Union nor the European Commission nor SERI can be held responsible for them. This research was also supported by a grant from the Simons Foundation (1013657, JL). R.K. is supported by the Helmholtz Association under the joint
research school "Munich School for Data Science - MUDS

The authors thank Matt Landreman for providing the strongly shaped boundary in Figure  \ref{fig:StarngeShape1} that started this work. 

%% file: Sections/Data.tex
\section*{Data Availability Statement}
The data that support the findings of this study are available from the corresponding author
upon reasonable request.

%% file: Sections/Interests.tex
\section*{Declaration of interest}
The authors report no conflict of interest.

%% file: Sections/Appendix.tex
\section{Derivation of EL equations}\label{Appendix1}
First, we take the variation of the term proportional to the Jacobian squared, leading to 
\begin{align}
         \delta \integral\, \frac{1}{2} f \sqrt{g}^2  & = \integral\left[ \textstyle f  \displaystyle \sqrt{g} \delta \sqrt{g} + \frac{\sqrt{g}^2}{2} \frac{\delta f}{\delta \textbf{x}} \cdot \delta \textbf{x}\right] \  \\
         &  =  - \integral  \sqrt{g}\delta \textbf{x}\cdot\Big[ \nabla (f \sqrt{g}) - \frac{\sqrt{g}}{2} \frac{\delta f}{\delta \textbf{x}}\, + \notag\\
         &  \quad \quad  f\left(\frac{\partial}{\partial s} (\sqrt{g} \nabla s) + \frac{\partial}{\partial \theta}(\sqrt{g}\nabla \theta) + \frac{\partial}{\partial \zeta} (\sqrt{g} \nabla \zeta)\right) \Big] \label{Eq:bho}\\ 
         & =  - \integral \sqrt{g}\delta \textbf{x}\cdot\left[ \nabla (f \sqrt{g}) - \frac{\sqrt{g}}{2} \frac{\delta f}{\delta \textbf{x}} \right].
\end{align}
$\delta f / \delta \textbf{x}$ is the variation of $f$ with respect to the mapping $\textbf{x}$. From the first to the second line, we used the definition of $\sqrt{g}$ and integrated by parts.
In Appendix \ref{sec:AppendixA}, the explicit application of the boundary conditions in the integration by parts is presented. $\textbf{x}$ being analytical and $s f$ being finite in the limit of approaching the coordinate axis are sufficient conditions for vanishing the local boundary terms. As shown in  Appendix \ref{sec:AppendixB},  the last term in Eq. (\ref{Eq:bho}) is zero using the Christoffel symbols \citep{wald2010general}. 

Regarding the variation in the radial length:
\begin{align} \label{Eq:radialcurvature}
   \integral \sqrt{\textbf{e}_s \cdot \textbf{e}_s} = & \integral \frac{\textbf{e}_s}{|\textbf{e}_s|} \cdot \pdv{\delta \textbf{x}}{s} =  - \integral \delta \mathbf{x} \cdot \boldsymbol{\kappa},
\end{align}
with $\boldsymbol{\kappa} = (\textbf{e}_s \cdot \nabla) \textbf{e}_s$. As illustrated in Appendix \ref{sec:AppendixC}, from analyticity at the coordinate axis the boundary terms from integration by parts vanish. 
The variation of the action in Equation (\ref{eq:Action}) with respect to the coordinate mapping yields:
\begin{equation}\label{EL}
\delta \mathcal{S} = - \integral \sqrt{g} \bigg[ \nabla (f \sqrt{g}) - \frac{\sqrt{g}}{2} \frac{\delta f}{\delta \textbf{x}} + \frac{\omega} {\sqrt{g}}\boldsymbol{\kappa} \bigg] \cdot \delta \textbf{x}.
\end{equation}

The stationary points of $\mathcal{S}$ are obtained when $\delta \mathcal{S}=0$ for arbitrary $\delta\textbf{x}$. 
Since Equation (\ref{EL}) holds for arbitrary $\delta \textbf{x}$ satisfying the boundary conditions, implying the vanishing of the expression inside the square brackets for a stationary point, the Euler-Lagrange equations for the coordinate mapping $\textbf{x}(s,\theta, \zeta)$ are derived.

\section{Boundary terms in the EL equations}\label{sec:AppendixA}
We expand the first term on the right side of Eq. (\ref{Eq:bho}).
Using the definition for $\sqrt{g}$:
\begin{equation}\label{eq:action}
   \integral f \sqrt{g}\left[\pdv{\delta \mathbf{x}}{s} \cdot(\textbf{e}_{\theta} \cross \textbf{e}_{\zeta}) + \pdv{\delta \mathbf{x}}{\theta} \cdot (\textbf{e}_{\zeta} \cross \textbf{e}_{s}) + \pdv{\delta \mathbf{x}}{\zeta} \cdot(\textbf{e}_{s} \cross \textbf{e}_{\theta})\right].
\end{equation}
The integration by parts in $s$ leads to:
\begin{equation}
    \int_0^1 \! \! \!ds \, f \sqrt{g} \pdv{\delta \mathbf{x}}{s} \cdot(\textbf{e}_{\theta} \cross \textbf{e}_{\zeta}) = f \sqrt{g}\delta \mathbf{x} \cdot(\textbf{e}_{\theta} \cross \textbf{e}_{\zeta}) \Bigg|_0^1 - \int_0^1\!\!\! ds \,\delta \mathbf{x} \cdot \frac{\partial}{\partial s} (f\sqrt{g} \textbf{e}_{\theta} \cross \textbf{e}_{\zeta}).
\end{equation}
The local term evaluated at the boundary is zero, since at $s = 1$ $\delta \mathbf{x}$ vanishes. The term at the axis gives a local contribution that needs to be evaluated in the limit where $s$ is zero:
\begin{equation}\label{eq:ultima}
    \int_0 ^{2\pi}\!\!\!\!\!d\zeta \int_{0}^{2\pi}\!\!\!\!\!d\theta\, f \sqrt{g}\delta \mathbf{x} \cdot(\textbf{e}_{\theta} \cross \textbf{e}_{\zeta}) \Big|_{s = 0}.
\end{equation} 
We prove the integrated part in Eq. (\ref{eq:ultima}) is zero for each $\delta \textbf{x}$. Since the mapping is analytical near the axis, we have that $\mathbf{e}_\theta  = 0$ and $\mathbf{e}_\zeta$, $\mathbf{e}_s$  are finite. We are left to prove that $f \sqrt{g}$ is finite in the limit of $s$ vanishing. This is equivalent to checking that $f \textbf{e}_\theta$ is a finite quantity. Using the condition of analyticity at the coordinate axis:
\begin{equation}\label{eq:expansion}
\begin{split}
     f \textbf{e}_\theta = & f \frac{\partial}{\partial \theta}  \left(\textbf{x}_a(\zeta) + \textbf{e}_s \Big|_{s = 0} s + \pdv[2]{\textbf{x}}{s}\Big|_{s = 0} s^2 + \mathcal{O}(s^3)\right) \\
     = & f s \pdv{\textbf{e}_s}{\theta}\Big|_{s = 0} +  f  \frac{\partial}{\partial \theta} \pdv[2]{\textbf{x}}{s}\Big|_{s = 0}s^2 + \mathcal{O}(s^3),
\end{split}
\end{equation}
where $\partial_s \textbf{x}_a = 0$ and with $\partial_{s}^n \textbf{x}$ and its derivatives to $\theta$ are well defined. Eq (\ref{eq:expansion}) states that for $f\textbf{e}_\theta$ to be defined for $s$ going to zero, we need $f s$ to either be fixed or go to zero simultaneously. We have that the term in (\ref{eq:ultima}) is zero in these conditions for every $\delta \textbf{x}$.

Going back to Eq. (\ref{eq:action}), integrating by parts along the poloidal angle $\partial_\theta \delta \mathbf{x} \cdot (\mathbf{e}_\theta \cross \textbf{e}_\zeta)$:
\begin{equation}
    \int_0^{2 \pi} d\theta \, f \sqrt{g}\pdv{\delta \mathbf{x}}{\theta} \cdot(\textbf{e}_{\zeta} \cross \textbf{e}_{s}) = f \sqrt{g}\delta \mathbf{x} \cdot(\textbf{e}_{\theta} \cross \textbf{e}_{\zeta}) \Bigg|_0^{2 \pi} - \int_0^{2\pi} d\theta \, \delta \mathbf{x} \cdot\frac{\partial}{\partial \theta}(f \sqrt{g}\textbf{e}_{\zeta} \cross \textbf{e}_{s}).
\end{equation}
Due to periodicity, the local term is zero. The same steps are applied for the integration over $\zeta$.

\section{Useful identities for EL derivation} \label{sec:AppendixB}
We now prove that the third term on the right side of Eq.(\ref{Eq:bho}) vanishes. To light up the notation and be clearer in the computations, we introduce $\xi^i$, with $i = 1, 2, 3$ as $(s, \theta, \zeta)$, and $x_i$ are the Cartesian coordinates for the mapping. We use Einstein's notations, where the sum over repeated indices is omitted.
\begin{align}
    \frac{\partial}{\partial \xi^i}\big(\sqrt{g} \nabla \xi^i\big) = \nabla \sqrt{g} + \sqrt{g} \pdv{\nabla \xi^i}{\xi^i}.
\end{align}
Considering the second term on the right side, 
\begin{align}
 \pdv{\nabla \xi^i}{\xi^i} = \pdv{x_m}{\xi^i} \pdv{\nabla \xi^i}{x_m}.
\end{align}
Using the identity \citep{wald2010general}, 
\begin{align}
   \pdv{\nabla \xi ^i}{x^j} = - \Gamma^{i}_{j k}\nabla \xi^k,
\end{align}
\begin{align} \label{eq:Cri2}
     \pdv{\nabla \xi^i}{\xi^i} = - \delta_i^k \pdv{x_m}{\xi^k} \Gamma_{m n}^i \nabla \xi ^n.
\end{align}
From the Christoffel identity $\Gamma^i_{ik} = \partial_k \log \sqrt{g}$, then
\begin{align} \label{eq:Cri1}
    \pdv{x_k}{\xi_m} \Gamma_{jk}^i \delta^j_i = \pdv{\log \sqrt{g}}{\xi^m}.
\end{align}
Substitute Eq. \ref{eq:Cri1} into Eq. \ref{eq:Cri2}, we obtain that:
\begin{align}
     \pdv{\nabla \xi^i}{\xi^i} = - \pdv{\log \sqrt{g}}{\xi^i} \nabla \xi^i = - \nabla \log \sqrt{g}.
\end{align}
Giving,
\begin{align}
    \frac{\partial}{\partial\xi^i }\big(\sqrt{g} \nabla \xi^i\big) = \nabla \sqrt{g} - \sqrt{g} \nabla\log \sqrt{g} = 0.
\end{align}

\section{Local curvature term} \label{sec:AppendixC}

The local term in $s$ obtained by integrating by parts the right side of Eq. (\ref{Eq:radialcurvature}) is:
\begin{equation}\label{eq:firstlocal}
    \int_0^{2\pi}\!\!\!\!\! d\theta \int_0^{2\pi}\!\!\!\!\! d \zeta \, \frac{\textbf{e}_s}{|\textbf{e}_s|}\cdot \delta \mathbf{x} \Big|_{s = 0}.
\end{equation}
Writing it explicitly in the limit of vanishing $s$:
\begin{equation}\label{eq:integral}
    \int_0^{2\pi}\!\!\!\!\! d\theta \int_0^{2\pi}\!\!\!\!\!d\zeta \,  \frac{\delta x_R \sum_n \Tilde{x}_n \cos(\theta - n \zeta)\mathbf{R} + \delta x_Z \sum_n \Tilde{y}_n \sin(\theta - n\zeta) \mathbf{Z}}{\sqrt{(\sum_n\Tilde{x}_n \cos(\theta - n \zeta))^2 + (\sum_n\Tilde{y}_n \sin(\theta - n\zeta))^2}},
\end{equation}
and using the Ptolemy’s identities, the integral in Eq (\ref{eq:integral}) can be rearranged into:
\begin{equation}
\begin{split}
       & \int_0^{2\pi}\!\!\!\!\! d\theta \int_0^{2\pi}\!\!\!\!\!d\zeta \, \frac{ \delta x_R(a\cos(\theta) + b \sin(\theta))\mathbf{R} + \delta x_Z( c\cos(\theta) + d\sin(\theta))\mathbf{Z}}{\sqrt{(a\cos(\theta) + b \sin(\theta))^2 + ( c\cos(\theta) + d\sin(\theta))^2}}, \\
       & a = \sum_n \Tilde{x}_n\cos(-n\zeta), \\
       & b = \sum_n \Tilde{x}_n\sin(-n\zeta), \\
       & c = - \sum_n \Tilde{y}_n\sin(-n\zeta), \\
       & d = \sum_n\Tilde{y}_n\cos(-n\zeta).\\
\end{split}
\end{equation}
Since $a, b, c, d$ are functions only of $\zeta$, the integration along $\theta$ from $0$ to $2 \pi$ for each component is a tabulated integral whose value is zero \citep{jeffrey2008handbook} for every choice of $\delta \textbf{x}$.

\section{Boundary coefficients for convergence study} \label{ultima}

For the axisymmetric case, the boundary coefficients in Figure \ref{fig:ELright} are $R_0^b = 1$, $R_1^b = 0.6$, $R_2^b = 0.8$, $Z_0^b = 0$, $Z_1^b = 0.8$, and $Z_2^b = 0.1$, with $N_{s} \times N_{\theta} = 1000 \times 1000$. The boundary coefficients for the non-axisymmetric domain are in Table 1, where computations are performed with a grid of $N_{s} \times N_{\theta} \times N_{z} = 120 \times 120 \times 120$.
\begin{table}
\begin{center}
\begin{tabular}{l c c c} 
 \vspace{0.2cm}
 $n$ & $m$ & $R_{nm}^b$ & $Z_{nm}^b$ \\
 -1 & 0 & 0 & 0 \\ 
 0 & 0 & 1.0 & 0 \\
 1 & 0 & 0.2 &  0.4\\
 -1 & 1 & -0.2 & 0\\
 0 & 1 &  0.8 &  0.8\\ 
 1 & 1 &   0.03&  0.1\\ 
  -1 & 2 & 0.2&  0.4\\
 0 & 2 &  0.03&  0.03\\ 
 1 & 2 &  0.01&  0.05\\
 \label{tab:1}
\end{tabular}
\caption{The set of boundaries coefficients $R_{nm}^b$ and $Z_{nm}^b$ with $N = 1$ and $M = 2$.}
\end{center}
\end{table}

\section{Gradient of $\mathcal{S}$ in Zernike-Fourier space} \label{sec:AppendixD}

The action gradient is given as:
\begin{align}
    \pdv{\mathcal{S}}{X_{pqr}} = \sum_{ijk} \frac{1}{s_i}\mathcal{A}_{ijk} \pdv{\mathcal{A}_{ijk}}{X_{pqr}} + \omega \pdv{\mathcal{L}_{ijk}}{X_{pqr}},
\end{align}
with $X_{pqr} = r_{pqr}, \, z_{pqr}$. Its derivatives are given by, 
\begin{align}
    \pdv{A_{ijk}}{r_{pqr}} = &\frac{\text{sign}\big(\text{arg}(\mathcal{A}_{ijk})\big)}{2}\Bigg[ (Z_{i j + 1 k} - Z_{i + 1 j k}) \Bigg(\pdv{R_{ijk}}{r_{pqr}} - \pdv{R_{i + 1 j+ 1 k}}{r_{pqr}}\Bigg) + \\
    &+  (Z_{ijk} - Z_{i + 1j + 1k})\Bigg(\pdv{R_{ij + 1k}}{R_{pqr}} - \pdv{R_{i + 1jk}}{r_{pqr}}\Bigg)\Bigg], \nonumber \\
    \pdv{A_{ijk}}{z_{pqr}} & = 
    \frac{\text{sign}\big(\text{arg}(\mathcal{A}_{ijk})\big)}{2}\Bigg[ \Bigg( \pdv{Z_{ij + 1k}}{z_{pqr}} - \pdv{Z_{i + 1 j k}}{z_{pqr}}\Bigg) (R_{ijk} - \pdv{R_{i + 1j+ 1k}}{r_{pqr}}) + \\
    &+  \Bigg(\pdv{Z_{ijk}}{z_{pqr}} - \pdv{Z_{i + 1j + 1k}}{z_{pqr}}\Bigg)(R_{ij + 1k} - R_{i + 1jk})\Bigg] \nonumber,\\
    \pdv{\mathcal{L}_{ijk}}{r_{pqr}} & = \frac{R_{i+1jk} - R_{ijk}}{\mathcal{L}_{ijk}} \Bigg(\pdv{R_{i+1jk}}{r_{pqr}} - \pdv{R_{ijk}}{r_{pqr}}\Bigg), \\
    \pdv{\mathcal{L}_{ijk}}{z_{pqr}} & = \frac{Z_{i+1jk} - Z_{ijk}}{\mathcal{L}_{ijk}} \Bigg(\pdv{Z_{i+1jk}}{z_{pqr}} - \pdv{Z_{ijk}}{z_{pqr}}\Bigg), \\
    \pdv{R_{ijk}}{r_{pqr}} & = \cos(r \theta_j - p N_{fp} \zeta_k)\Big(\mathcal{Z}^q_r(s_i) - \mathcal{Z}^r_r(s_i)\Theta(q - r - 1)\Big) \\
       \pdv{Z_{ijk}}{z_{pqr}} & = \sin(r \theta_j - p N_{fp} \zeta_k)\Big(\mathcal{Z}^q_r(s_i) - \mathcal{Z}^r_r(s_i)\Theta(q - r - 1)\Big),
\end{align}
with $\Theta(x) = 0$ if $x < 0$ and $\Theta(x) = 1$ otherwise.